\documentclass[review]{elsarticle}

\usepackage{lineno,hyperref}
\modulolinenumbers[5]

\usepackage{graphicx}
\usepackage{dcolumn}
\usepackage{bm}
\usepackage{multirow}
\usepackage{textcomp}
\usepackage[]{subfigure}

\usepackage{booktabs}
\usepackage{array}
\usepackage{pgf}
\usepackage{graphicx}
\usepackage[absolute,overlay]{textpos}
\usepackage{calc}
\usepackage{amsmath}
\usepackage{amsfonts}
\usepackage{amssymb}
\usepackage{textcomp}
\usepackage{tikz}
\usepackage{gensymb}
\usepackage{color,soul}
\usepackage{xcolor}

\newcommand{\commentCWS}[1]%
{\textsf{\textcolor{blue}{#1$^{\mathrm{CWS}}$}}}
\usepackage[colorinlistoftodos]{todonotes} 

\newcommand{\figwidth}{0.9} 

\begin{document}

\begin{frontmatter}

\title{Predicting the effect of strain path on the strain aging behaviour of Ultra-low carbon steel}


\author[ubc]{W.~A.~Melo}
\author[ubc]{C.~W.~Sinclair\corref{cor1}}
    \ead{chad.sinclair@ubc.ca}

\cortext[cor1]{Corresponding Author,}

\address[ubc]{Department of Materials Engineering, The University of British Columbia, 309-6350 Stores Road, Vancouver, Canada} 

\begin{abstract}
A recently presented model is adapted to predict the effects of strain path change on the strain aging (bake hardening) behaviour of an ultra low carbon (ULC) steel.  Samples pre-deformed by rolling were aged and then tested in uniaxial tension at 0, 45 and 90$^\circ$ degrees to the prior rolling direction.  The results show that aging changes not only the yield strength of the material but also the work hardening rate.  The increase in yield strength is interpreted to be dominated by a reduction in mobile dislocation density, this reflecting the classic ideas of Hahn and others. The change in work hardening behaviour is treated as arising from the different role played by solute `loaded' dislocations compared to dislocations generated by straining after aging.

\end{abstract}


\end{frontmatter}

\date{\today}


\section{Introduction} 

Ultra-low carbon (ULC) steels rely primarily on two phenomena to achieve strength in formed parts; work hardening and strain aging.  These two features work in tandem in the bake hardening process; steel is first deformed, to work harden it, followed by annealing to allow for a further increase in strength due to interstitial solute re-arrangement and segregation to dislocations (strain aging or bake hardening) (see e.g. \cite{Ballarin2009a, Pereloma2017}).  Laboratory characterization of strain aging in ULC steels is conventionally carried out by performing tensile pre-straining ($\sim$2-5\% tensile strain, consistent with the strain imposed during stamping operations) then aging at 150-200$^\circ$C for 20-30 minutes to simulate the industrial paint bake cycle \cite{Ballarin2009a}.  This aged material is then tested again in tension, the orientation of the tensile axis being kept the same between tests. The typical result of this process, highlighted in Figure \ref{fig:strainagekinetics}a, is an increase in the yield strength post strain aging, with the appearance of an upper yield point and L\"uders plateau.  
Commercially, bake hardening (BH) is quantified as the increases in yield stress following the aging treatment. This definition uses the difference between the flow stress at the end of pre-deformation and the upper or lower yield stress measured after aging.  The aging time-temperature dependence of BH has been studied in detail with a number of analytical models that allow for a phenomenological description of this behaviour, see e.g. \cite{A.K.De.2004,Ballarin2009a,De2000,Berbenni2004}.  
Despite this convention for the measurement of BH, the strain path experienced by steel components during fabrication and service are usually neither uniaxial tension nor performed in the same direction pre and post aging.   For example, in sheet metal forming for automotive applications the pre-deformation typically ranges from plane strain to equibiaxial tension, while the deformation of interest post aging, e.g. ‘denting’, is also far from simple uniaxial tension \cite{Ballarin2009a,Ballarin2009b}.  The effect of strain path change following strain aging has been most heavily studied under conditions where the strain path is reversed between the pre-deformation and the deformation post aging.  Without aging, such tests (e.g. tension-compression or torsion-reverse torsion) are already complicated by the Bauschinger effect \cite{Saada2013}. For pre-strain levels commensurate with commercial bake hardening (a few percent), anisotropy of mechanical response on strain path change is most often (though not always, see e.g. \cite{Vincze2005}) considered to arise due to low misorientation dislocation cells formed in the first stage of deformation \cite{Wilson1994,Schmitt1991,Beyerlein2007,Bate2007}. These cells can act as strong barriers to the operation of (new) slip systems on reloading, though typically the material evolves quickly back towards the work hardening response it would have had if it had been deformed monotonically \cite{Wilson1994,Beyerlein2007}. In work on steel without strain aging, Wilson and Bate \cite{Wilson1994} showed that yielding upon load path change was accompanied (microscopically) by the inhomogeneous operation of slip systems in some grains first, leading to localized slip bands. Continued deformation then led to plasticity spreading through the sample.

When strain aging is performed between deformation steps where a strain path change is involved, the most common observation, for a range of steels, has been that strain aging acts to reduce, though never totally eliminate, the Bauschinger effect \cite{Sowerby1977,Elliot2004,D.N.Williams1980,Lai2018}.  This means that, unlike tension-tension tests where the BH effect is characterized by an upper and lower yield point following aging, in strain path reversal tests the stress-strain curve remains smooth and the deformation remains uniform within the gauge section of the sample.  Examples of non-reversed strain path changes with strain aging (e.g. tension-aging-tension at an angle to the original tensile direction) are rare in the literature and the results less clear concerning the impact on the mechanical response.  For example, Jun et al. \cite{Jun1986} reported that the yield strength increases in tension-tension tests, regardless of the angle between the first and second directions of loading, while Hiwatashi et al. \cite{Hiwatashi2009} reported both increase and decrease in yield strength.  

What is consistent among all studies where strain aging is conducted with a change in strain path is that the strong yield point phenomena characteristic of strain aged samples subjected to monotonic tension-tension tests is not present.  However, aging conducted with a strain path change does result in a significant change in the work hardening behaviour.  This was noted by Ballarin et al. \cite{Ballarin2009a,Ballarin2009b} who developed a phenomenological model to predict not only the yielding but also the work hardening response of samples subjected to strain aging and non-uniaxial, non-monotonic strain paths.  The model, while able to reproduce the complex features of the strain aging response observed under a variety of strain path changes, relies on a large number of empirical parameters which need to be adjusted for each condition, this limiting the model’s predictive capabilities.
The work presented below seeks to build from the work of Ballarin et al. \cite{Ballarin2009b} by proposing an alternative model for predicting the strain aging response of ULC steel samples subjected to a strain path change.  The physically-based model used here is adapted from recent work which itself is strongly motivated by the classic work of Hahn \cite{Hahn1962}.  Experiments are presented both for conventional tension-tension strain aging tests as well as for tests where samples were pre-deformed in cold rolling, aged, then tested in tension.  The strengths and weaknesses of this approach are highlighted as are areas for potential future work.

\section{Materials and Experimental Method}

For this study a 1~mm thick ULC steel sheet (composition in Table \ref{tab:comp}) was supplied by ArcelorMittal Dofasco. The material, received in the as-cold rolled state, was annealed in a box furnace in air at 800$^\circ$C for 7 min, the temperature of the samples monitored by a thermocouple spot welded to them. Annealing resulted in a fully recrystallized microstructure with an average equal area grain diameter of 14 $\mu$m.

\begin{table*}[htbp]
\resizebox{\textwidth}{!}{%
 \centering
 \begin{tabular}{cccccccccccccc}
  \toprule
   & \textbf{C} & \textbf{N} & \textbf{Si} & \textbf{P} & \textbf{Mn} & \textbf{Nb} & \textbf{Ti} & \textbf{S} & \textbf{Al (tot)} & \textbf{Cr} & \textbf{Ni} & \textbf{Cu} &\textbf{Other}  \\
  \midrule
			& 0.0024 & 0.0027 & 0.011 & 0.039 & 0.358 & 0.0048 & 0.012 & 0.0084 & 0.042 & 0.043 & 0.023 & 0.037 & 0.006 \\
			
  \bottomrule  
 \end{tabular}}
 \caption{Nominal composition of studied ULC steel.  Balance is iron.}
 \label{tab:comp}
\end{table*}

 In order to study strain aging on this material, two forms of pre-deformation were used; uniaxial tensile tests were performed on one set of samples, while another set of samples were pre-deformed by performing laboratory cold rolling.  In both cases, following aging, the second deformation (to measure the strain aging effect) was performed in uniaxial tension.  
For all tensile tests the same dog-bone shaped tensile samples were used (gauge length of 40 mm, gauge width of 6 mm).  These were tested at room temperature in an Instron load frame with a fixed cross-head speed of 0.21 mm/s, corresponding to a nominal strain rate of approximately $6 \times 10^{-2}$ s$^{-1}$.  Strains were measured using a clip-on extensometer.  For the tensile tests performed to pre-strain the material, the tests were stopped at 10\% total strain, this being just beyond the end of the L\"uder’s plateau in the stress strain curves.  

For samples pre-deformed by rolling, the deformation was conducted using an unlubricated laboratory rolling mill.  The rolling direction was kept the same as that used in the commercial production of the sheet.  All sheets were cold rolled in 4-5 passes to a reduction in thickness of 9 $\pm$ 1\%.  The reduction in thickness was measured using a digital micrometer at several locations in the as-rolled sheet.  From these rolled sheets, tensile samples were cut parallel (0$^\circ$) to the sheet’s rolling direction (RD), transverse  (90$^\circ$) to RD, and at 45$^\circ$ to RD.  Tensile samples cut at 0 and 90$^\circ$ to RD were processed in two sheets rolled at the same time, meaning that the same reduction per pass and total reduction were achieved.  Samples cut at 45$^\circ$ to RD were taken from a sheet rolled separately.

Following pre-deformation in tension or in rolling, samples were aged in a stirred silicone oil bath.  The bath temperature was monitored using a K-type thermocouple submerged next to the samples. For the samples pre-deformed in tension, the aging treatment was performed within 10 minutes of the end of the pre-deformation step.  For samples pre-deformed in rolling, a longer delay between deformation and aging was required owing to the need to machine tensile samples.  In this case, aging was limited to long times/high temperatures so as to minimize the effects of further aging at room temperature.  

Two-dimensional digital image correlation (DIC) was performed on selected samples during tensile testing to clarify the homogeneity of strain in the gauge section.  For DIC measurements a single camera was used with the LAVISION DIC system for image capture and subsequent processing.  A random speckle pattern was deposited on the surface of these samples using black paint and an airbrush.  During DIC image acquisition the measured load from the load cell was recorded on each image.  This allowed for a direct correlation between the DIC measurements and the stress-strain response.  As a validation of the technique, the average axial strain in the gauge section of the sample was used, along with the recorded load, to calculate the Young’s modulus of samples during unloading.  The results were found to always fall between 204-208 GPa. 

Finally, to assist with the prediction of the stress-strain response under conditions where non-uniform strains were observed during tensile testing (due to L\"uders bands), simulations of tensile tests were performed using ABAQUS CAE version 6.14 FEA software \cite{abaqus}.  The full tensile sample geometry was used for these simulations, with linear 8-node brick elements (C3D8R). Elements within the gauge section were 0.245 mm wide, 0.3 mm long and 1 mm in thick (1 element in the thickness of the sample). An encastre boundary condition (no displacement or rotation) was imposed on the nodes in the lower grip section, while nodes on the upper grip surface were subjected to a displacement parallel to the tensile axis, with no displacements/rotations allowed in any other directions.  The material in the gauge section was subjected to no mechanical constraint. To allow for the controlled nucleation of a L\"uders band, a very small geometric imperfection (similar to a through thickness v-shaped notch) 90 $\mu$m deep and 20 $\mu$m wide was introduced at the junction between the top-left fillet and gauge section of the sample. 

 \section{Strain Aging Kinetics with a Monotonic Strain Path}
 
\begin{figure}[htbp]
\centering
\includegraphics[width=\figwidth\textwidth]{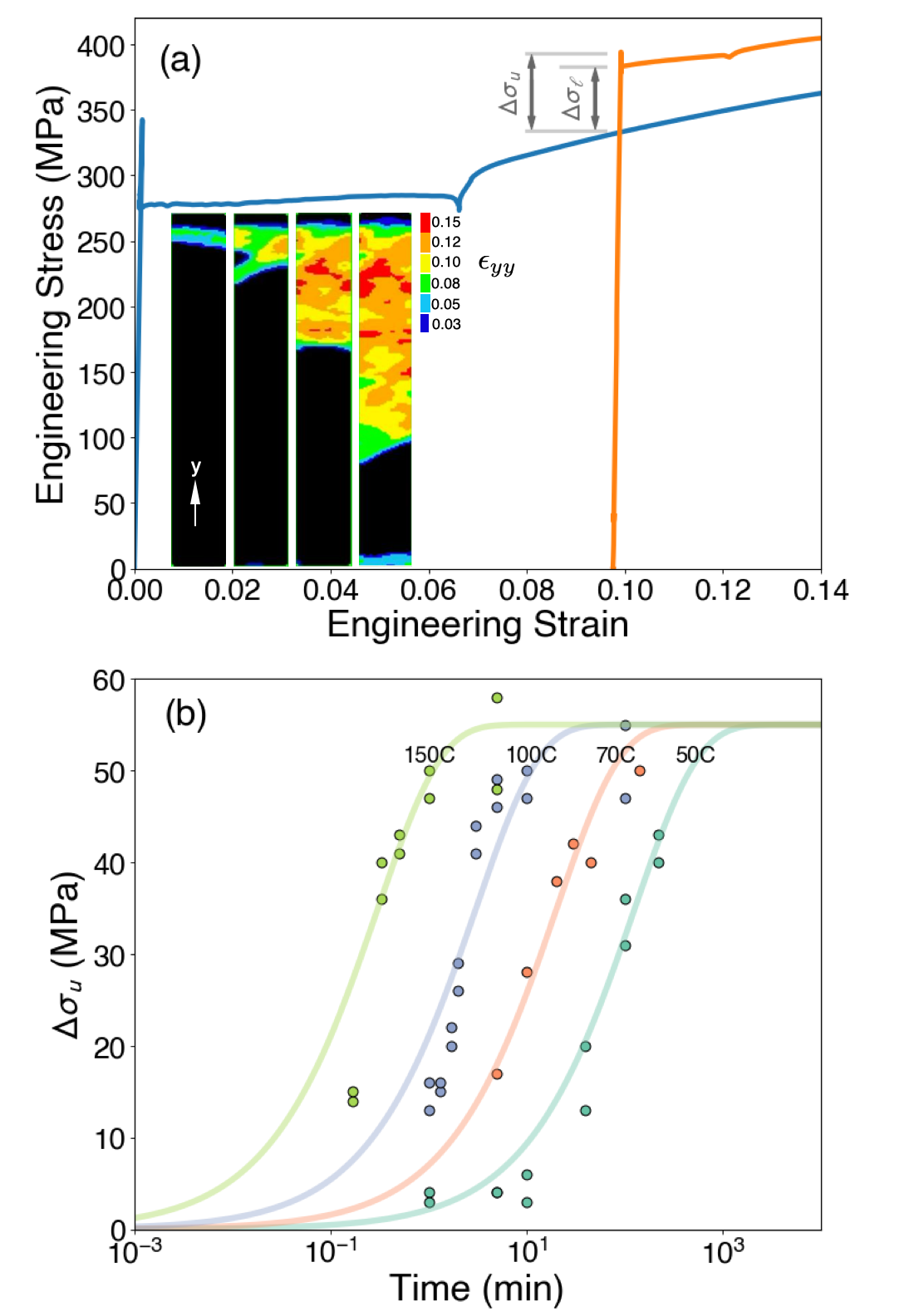}
\caption{a) Stress-strain curve of as-annealed material (blue) and strain aged material (10~min at 150$^\circ$C after 10\% tensile pre-deformation). Also shown for illustration are DIC images taken while the samples were undergoing L\"uders band propagation for the as-annealed material. b) strain aging kinetics for different times and temperatures following a 10\% pre-deformation in tension.  The solid lines show the best fits of equation \ref{eqn:agingkinetics}. }
\label{fig:strainagekinetics}
\end{figure}

 Strain aging tests were first performed under conventional tension-aging-tension conditions to assess the aging kinetics under the conditions used here.  Aging was conducted at 50, 100 and 150$^\circ$C for times ranging from 1 min to 220 min.  Figure \ref{fig:strainagekinetics}a shows the stress-strain curve for an unaged, or `as-annealed', tensile sample as well as a sample aged for 10 min at 150$^\circ$C, this illustrating the typical changes in stress-strain response following aging.  Also shown in the inset is the DIC measured propagation of the L\"uders bands in the as-annealed sample, the bands originating at the top and bottom of the sample’s gauge section.  Uniform deformation commences once these bands fill the gauge section of the sample. Figure \ref{fig:strainagekinetics}b summarizes the strain aging kinetics for all conditions studied, using the change in the upper yield stress ($\sigma_U$) as a measure of the strain aging/BH effect.  The solid lines in these plots represent a fit of the data to the classic Louat-Cottrell-Bilby model \cite{N.Louat1981,Cottrell1953},

\begin{equation}
\frac{\Delta \sigma_u}{\Delta \sigma_{max}} = 1-\exp{\left(- \left(\frac{t}{t^{*}}\right)^{2/3}\right)}
\label{eqn:agingkinetics}
\end{equation}

Using the data here it was found that a good prediction could be obtained using $\Delta \sigma_{max} = 55$ MPa and $t^*=2.34 \times 10^{-16} / D$ where $D = D_0\exp{\left(-\frac{Q}{RT}\right)}$ is the the diffusivity of carbon in $\alpha$-Fe, with  $D_0 = 1.2\times10^{-4}$ m$^2$/min and $Q = 85$ kJ/mol, as used in \cite{De2001}.  The results shown in Figure \ref{fig:strainagekinetics} agree well with previous results published in the literature for similar grades of steel, tested under similar circumstances.  While others have pointed to the possibility of additional hardening coming from the precipitation of carbides at higher temperatures/longer aging times, here the data seems to be adequately captured by the single set of kinetics represented by equation \ref{eqn:agingkinetics}. 

\section{Strain Aging with a Change in Strain Path}

Based on the above results, an aging condition of 10 min at 150$^\circ$C was selected for the rolled samples, as it represented the maximum BH effect observed in figure \ref{fig:strainagekinetics}.  As noted above, this `peak' strain aged condition was selected so as to minimize the affect of the samples needing to be left at room temperature for several hours prior to aging so that tensile samples could be machined. 

Like with the results shown in figure \ref{fig:strainagekinetics}, one would like to compare the mechanical (tensile) response after aging with the mechanical response of the material in the as-recrystallized state.  The challenge, in these experiments, arises from the ambiguity associated with defining an appropriate measure of equivalent plastic tensile strain to associate with the rolling pre-deformation.  Assuming the material obeys a Von Mises (J2) flow theory, one can estimate the equivalent plastic strain based on the thickness reduction ($\epsilon_t$) assuming monotonic plane strain deformation,

\begin{equation}
    \bar{\epsilon} = \frac{2}{\sqrt{3}}\epsilon_t
\end{equation}

This, however, should be considered as a lower bound estimate as it ignores the complex deformation path experienced as the material passes through the roll gap, including the effects of friction (see e.g. \cite{Lee1991,Hosford2011}).

\begin{figure}[htbp]
\centering
\includegraphics[width=\figwidth\textwidth]{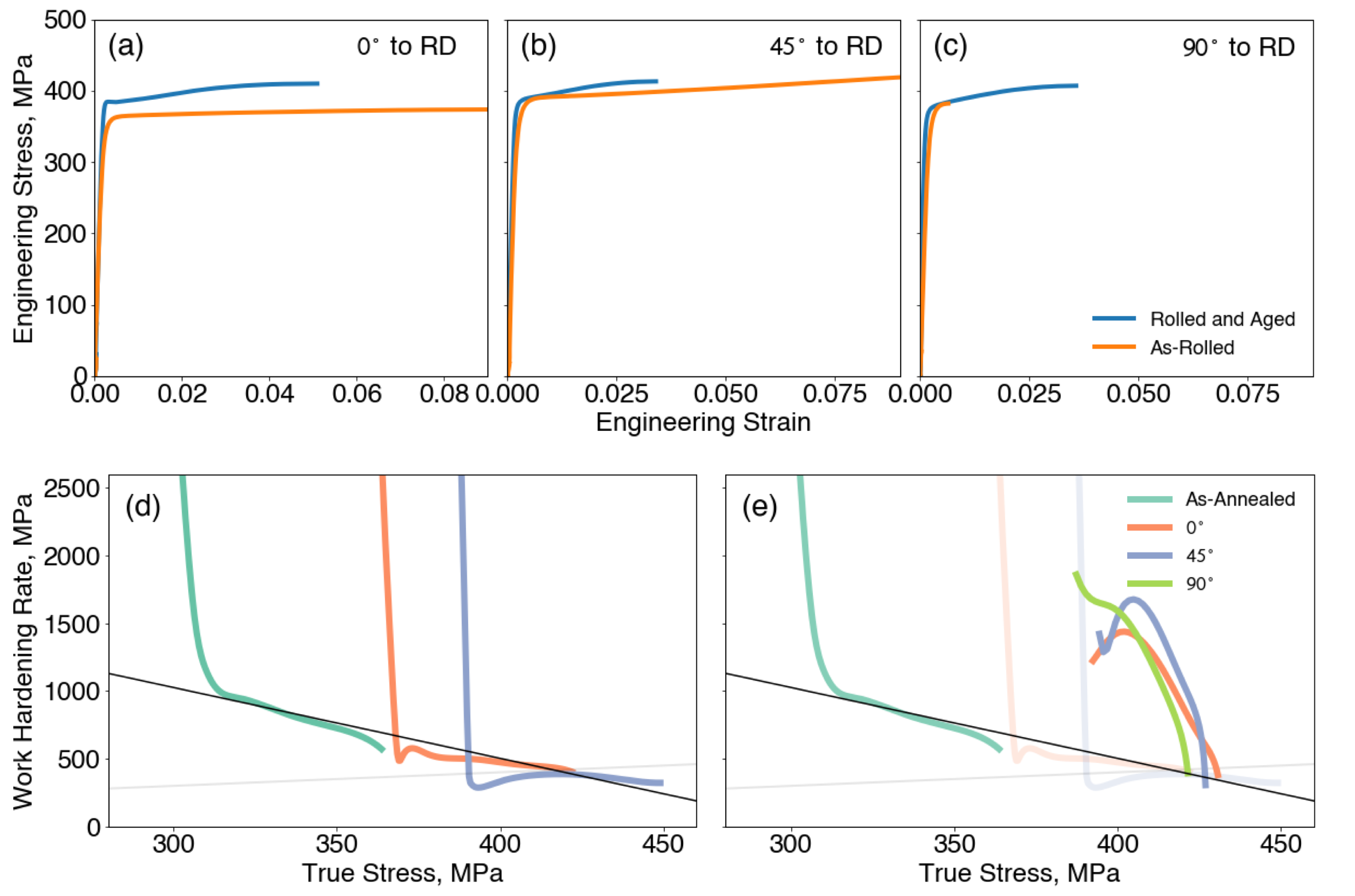}
\caption{a)-c) Stress-strain curves for samples after pre-deformation and pre-deformation and aging for samples rolled then tensile tested at 0$^\circ$, 45$^\circ$ and 90$^\circ$ to the rolling direction.  The aged samples were treated at 150$^\circ$C for 10 min. d) work hardening versus flow stress plot comparing the work hardening response (post L\"udering) to the work hardening response of the rolled and tensile tested samples e) as (d) but for samples that were aged after rolling.}
\label{fig:workhardeningafteraging}
\end{figure}

Additional data which can help reduce this ambiguity comes if we compare the work-hardening/flow stress relationship (Kocks-Mecking plot \cite{Kocks2003}) for samples with and without pre-deformation in rolling.  It is expected that the change in strain path may introduce a transient in the work hardening response \cite{VanRiel2007,Wen2016,Nesterova2001}, but that beyond this transient the work hardening-flow stress response, dominated by the dislocation density in this case, will reflect the underlying hardening response of the sample \cite{Mecking1981,Kocks2003}.  Figure \ref{fig:workhardeningafteraging}a-c shows the engineering stress-strain response of the as-rolled and rolled and aged samples tested at 0$^\circ$, 45$^\circ$ and 90$^\circ$ to the prior rolling direction.  In each case, the stress-strain curves are plotted to the onset of necking (plotted to the ultimate tensile strength).  As can be seen, in the case of the sample tested in the as-rolled condition at 90$^\circ$ to the tensile axis, necking commences shortly after yielding. Figure \ref{fig:workhardeningafteraging}d and e show the same data, but in this case as Kocks-Mecking plots where the work hardening behaviour of the starting as-annealed material (post L\"udering) is also shown.  Here the work hardening rates were computed from the true-stress, true strain data smoothed by interpolating using a B-spline with 5 knots and degree of 3 (as implimented in the splev function in scipy \cite{scipy}).  As conventional for cubic metals, the Kocks-Mecking plot for the as-annealed material shows a linear decrease in work hardening with increasing flow stress \cite{Mecking1981,Kocks2003}, the thin black line illustrates this for the three cases shown in figure \ref{fig:workhardeningafteraging}d.  The pre-deformed sample tested at 0$^\circ$ to RD shows a similar response and a work hardening behaviour that nearly overlaps (after an initial transient) with that of the as-recrystallized material.  This coincides with previous work which has shown that for steels, pre-deformed in rolling, small pre-strains in rolling followed by tensile testing parallel to RD gives the smallest strain-path change effect to the resulting stress-strain curve \cite{Wilson1994}.  In contrast, the rolled sample tested at 45$^\circ$ to RD shows a more significant transient in the work hardening response.  The sample tested at 90$^\circ$ to RD could not be analyzed owing to the early onset of localization. 

Figure \ref{fig:workhardeningafteraging}e shows that following aging, not only does the yield strength of the material change but so too does the work hardening behaviour.  In the case of all samples rolled, aged and tested at 0, 45 and 90$^\circ$ to RD, the initial work hardening increases as does the rate of drop of the work hardening rate with stress.  This observation would be unobtainable from a simple model that considers only changes in the yield strength of the material on aging (cf. equation \ref{eqn:agingkinetics}).

\begin{figure}[htbp]
\centering
\includegraphics[width=\figwidth\textwidth]{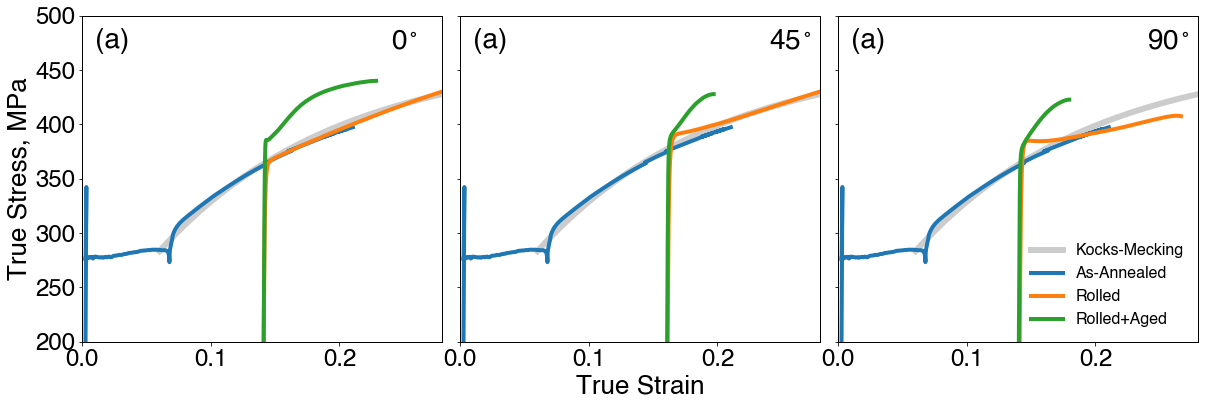}
\caption{a)-c) Stress-strain curves for samples after pre-deformation and pre-deformation and aging for samples rolled then tensile tested at 0$^\circ$, 45$^\circ$ and 90$^\circ$ to the rolling direction. Also shown are the stress-strain curves for the as annealed material.  The stress-strain curves for the rolled and rolled+aged samples have been translated in strain along the x-axis by $\bar{\epsilon} = 0.14$ for the samples tested at 0$^\circ$ and 90$^\circ$ to RD while the results for samples tested at 45$^\circ$ to RD were translated by $\bar{\epsilon} = 0.16$. The light line in the background (labeled `Kocks-Mecking' is the Kocks-Mecking model (equations \ref{eqn:taylor}, \ref{eqn:rhof1}) fit to the post-L\"udering portion of the as-annealed stress-strain curve.}
\label{fig:comparestressstrain}
\end{figure}

Following from the results in figure \ref{fig:workhardeningafteraging}, one can estimate an effective tensile strain resulting from rolling, this allowing us to adjust the initial strain used to plot the stress-strain curves in figure \ref{fig:workhardeningafteraging}a-c when comparing with the stress-strain curves of the as-annealed material.  From figure \ref{fig:workhardeningafteraging} we see that the work hardening behaviour of the samples rolled, then tested at 0 and 45$^\circ$ to RD follow a similar work hardening response to that of the as-annealed material, except for the transient observed in the behaviour of the sample tested at 45$^\circ$.  Using $\bar{\epsilon} > 0.1$ as a lower bound for estimating the equivalent strain introduced by rolling, we find that a good agreement between the stress-strain response of the as-annealed and rolled and tested samples can be obtained using $\bar{\epsilon} = 0.14$ for the samples tested at 0$^\circ$ to RD while the samples tested at 45$^\circ$ match well to a continuation of the stress-strain response of the as-annealed material with $\bar{\epsilon} = 0.16$ (figure \ref{fig:comparestressstrain}).  Of course, it is impossible to compare the stress-strain response of the as-rolled material tested at 90$^\circ$ to RD with the as-annealed material at higher strains as the ultimate tensile strength occurred just after yielding. In this case, however, since the samples taken at 0 and 90$^\circ$ to RD were cut from the same rolled sheet we have applied the same $\bar{\epsilon} = 0.14$ to both cases. 

\begin{figure}[htbp]
\centering
\includegraphics[width=\figwidth\textwidth]{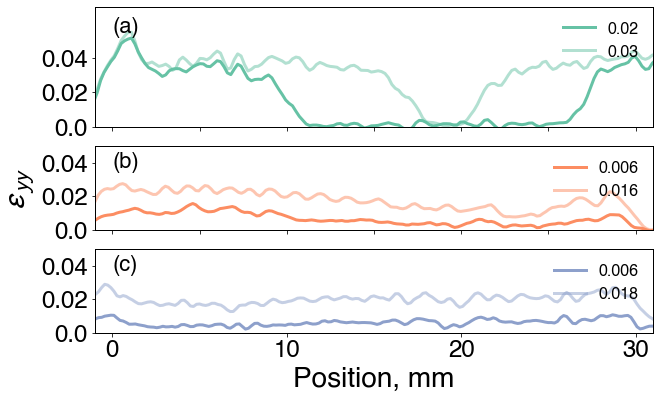}
\caption{Digitial image correlation measured tensile strain (parallel to the tensile axis) measured along the centre line of three tensile samples, the macroscopic strain for each curve is shown in the legends. a) Measurement made on a sample pre-deformed in tension, aged at 150$^\circ$C for 10 min then tested again in tension (same condition as the sample shown in \ref{fig:strainagekinetics}a). One can clearly see the propegation of L\"uders bands from both ends of the tensile sample toward the centre. b) Tensile samples tested after rolling and aging at 0$^\circ$ and c) 45$^\circ$ to RD.  While both the stress strain curves in these conditions showed some transient work hardening response on yielding, there is no evidence of strain localization suggesting L\"uders band formation or propegation.}
\label{fig:DIC}
\end{figure}

A final point regards the onset of yielding following rolling and aging.  While the classic response following strain aging is the one shown in figure \ref{fig:strainagekinetics}a, where after strain aging the material exhibits an upper and lower yield point accompanying a yield point elongation, the stress-strain curves shown in figure \ref{fig:workhardeningafteraging} appear quite different.  The samples rolled, aged then tested at 0 and 45$^\circ$ to RD both show some evidence of a low, transient hardening behaviour on yielding but this is not accompanied by any clear evidence of yield point elongation.  In the case of the sample rolled, aged and tested at 90$^\circ$ to RD, yielding gives way to a smooth work hardening response with no clear evidence of any transient.  To confirm that yielding occurred uniformly along the gauge length of these samples, DIC observations were made (figure \ref{fig:DIC}), these being compared to DIC measurements made on the gauge section of a sample pre-strained in tension, aged and then tested again in tension (cf. figure \ref{fig:strainagekinetics}b).  In this case we can see that the strain for the rolled and aged samples appears relatively uniform in the gauge section of the samples while the formation of L\"uders bands in the tensile pre-deformed and aged samples is clearly seen by the central portion of the sample having no measurable plastic strain.  This lack of L\"uders band formation in the rolled and aged samples greatly simplifies the analysis that will be presented next. 

\section{Predicting The Effect of Strain Aging Following a Strain Path Change}

As noted above, the classic approach to modelling the impact of strain aging on the mechanical response of ULC steel considers only the yield strength evolution.  Clearly this is insufficient to explain the changes shown in figure \ref{fig:workhardeningafteraging} where both yield strength and work hardening rate evolve.  A less used approach for explaining the effects of strain aging on steels was developed originally by Han \cite{Hahn1962}. In this approach, dislocations are considered to be `locked' by interstitial solute but that it is dislocation multiplication that controls the yielding behaviour, not dislocation `unlocking'.  One can take this approach a step further and consider the role that `locked' or solute `loaded' dislocations may have on the forest strengthening at higher strains.  This approach has recently been developed into a simplified two-state variable model (building from the original work of Kubin and Estrin \cite{Kubin1990}) and applied to Al-Mg alloys by Medrano \cite{Medrano2020}.  Here we test this approach to predict both the yield strength and work hardening response of the materials described above.  Rather than provide a detailed description of the model, we will focus on providing a basic overview, readers interested in a more detailed description are directed to \cite{Medrano2020}.  

The flow stress of the ULC steel is considered to be a function of the forest dislocation density ($\rho_f$) and the imposed strain rate ($\dot{\epsilon}$) according to,

\begin{equation}
    \sigma = \sigma_0 + M\alpha \mu b \sqrt{\rho_f}\left(\frac{\dot{\epsilon}}{\dot{\epsilon}_0}\right)^m
\label{eqn:taylor}
\end{equation}

In equation \ref{eqn:taylor}, $\sigma_0$ is taken to be the yield strength of the as-annealed material without the impact of strain aging, $M$ is the Taylor factor, $\alpha$ is the Saada constant. As-annealed samples tested at 0, 45 and 90$^\circ$ to RD were seen to have different (upper) yield stresses but the effect was small compared to those we hope to capture following aging, thus in this case a single value $M=3$ was used for all samples and no intrinsic anisotropy due to texture has been accounted for.  

The reference strain rate, $\dot{\epsilon}_0$, is assumed proportional to the mobile dislocation density $\dot{\epsilon}_0=\beta^*\rho_m$  \cite{Kubin1990}.  In  \cite{Medrano2020} a simple approximation is taken for the evolution of the mobile dislocation density, its value evolving towards a saturation value $\rho_m^{sat}$ which is taken as a parameter in the model.  This evolution is controlled by a parameter $\eta$ as,

\begin{equation}
    \frac{d\tilde{\rho}_m}{d\epsilon} = \eta\left(1-\tilde{\rho}_m\right)
\label{eqn:rhom}
\end{equation}

where $\tilde{\rho}_m = \rho_m/\rho_m^{sat}$. 

The evolution of the forest dislocation density, in the absence of aging, is taken to obey the classic Kocks-Mecking phenomenological evolution equation \cite{Mecking1981},

\begin{equation}
    \frac{d\rho_f}{d\epsilon} = k_1\sqrt{\rho_f}-k_2\rho_f
\label{eqn:rhof1}
\end{equation}

With strain aging, however, one needs to consider the role that solute `loaded' dislocations have on this equation for dislocation storage.  In this case, we additionally separate forest dislocations into two types, those which have been `loaded' with solute (density $\rho_{f}^*$) and those that are not ($\rho_f$).  As in \cite{Medrano2020} we allow for the fact that both types of forest dislocation will impact on the storage of additional forest dislocations adding an additional term to the above evolution equation,

\begin{equation}
    \frac{d\rho_f}{d\epsilon} = k_3\sqrt{\rho_{f}^*}+k_1\sqrt{\rho_f}-k_2\rho_f
\label{eqn:rhof2}
\end{equation}

In \cite{Medrano2020} an additional storage term due to mobile dislocations was included in equation \ref{eqn:rhof2}, consistent with \cite{Kubin1990}, but it has been found here that this term did not significantly contribute to improving the prediction (i.e. its pre-factor was small) and so it has not been included so as to reduce the number of parameters in the model.

Taking the above points into consideration, a single test on a solution treated sample, allows one to establish values for $\sigma_0$, $k_1$ and $k_2$, the values found for the alloy studied here being given in Table \ref{tab:params}.  As in \cite{Medrano2020} the density of `loaded' dislocations is allowed to be reduced by deformation, this obeying the simple law,

\begin{equation}
    \frac{d\rho_{f}^*}{d\epsilon}  =-k_4\rho_{f}^*\left[1-\exp{\left(\sqrt{\frac{\rho_{f}^*}{\rho_f}}\right)}\right]
\label{eqn:rhof*}
\end{equation}

The results of this are a model which requires the evaluation of $\sigma_0$, $k_1$, $k_2$, $k_3$, $k_4$, $m$, $\eta$, $\rho_m^{sat}$ and $\rho_{f}^*$ at the end of strain aging in order to predict the stress-strain response of a sample after deformation and aging.  While the number of parameters seems large we will show here that several of these must be held constant over all conditions considered and that others are relatively insensitive to model predictions.

We can first obtain estimates for $\sigma_0, k_1$ and $k_2$ by considering the behaviour of the as-annealed material.  Specifically we assume a condition where $\rho_m \approx \rho_m^{sat}$ and $\rho_{f}^* = 0$. This implies that rather than deal with fitting to the full stress-strain response in figure \ref{fig:strainagekinetics}a, which would require a spatially resolved model for the tensile sample to include the inhomogeneous strain and strain rate at the L\"uders band front, here we fit our response to the post-L\"uders portion of the stress-strain curve.  While the local strain rate (and therefore the evolved dislocation density) within the portion of the stress-strain curve exhibiting L\"udering will not be the same as that obtained from the same deformation performed uniformly, we consider this for now a second order effect, something we will return to at the end of the discussion.  The values for these three parameters were found from a best fit to the as-annealed work hardening curve shown in figure \ref{fig:workhardeningafteraging} and are reported in table \ref{tab:params}.  The resulting smooth stress-strain response is also shown in figure \ref{fig:workhardeningafteraging} (thin black line in d and e) and figure \ref{fig:comparestressstrain} (as the light grey lines in the background of each figure). 

With these parameters established, it is left to find $k_3,k_4, \eta, \rho_m^{sat}$ as well as the initial values (after rolling and/or rolling and aging) of $\rho_f$, $\rho_{f}^*$ and $\tilde{\rho}_{m}$.  The first four of these parameters are taken to be constant regardless of the testing conditions (i.e. same values are used for the as-rolled and the rolled and aged samples). In contrast, the values of the initial dislocation densities ($\rho_f$, $\rho_f^*$ and $\rho_m$) will be established for each of the test conditions as described below.  As in \cite{Medrano2020}, the model is relatively insensitive to the value of $\eta$ and $\rho_m^{sat}$, thus we have simply fixed these values to ones similar to those used in \cite{Medrano2020}.  These parameters mainly affect the initial yielding response as the mobile dislocation density quickly approaches $\rho_m^{sat}$ on deformation.  The values of $k_3$ and $k_4$ are particularly important for predicting the initially high work hardening rate seen in figure \ref{fig:workhardeningafteraging}e and its rapid drop with straining.  Eventually, as $\rho_{f}^*$ drops the work hardening-flow stress relationship predicted by the above equations are expected to asymptotically rejoin that of the as-annealed sample as the fraction of loaded dislocations is removed.

For the as-rolled samples, the initial values of the dislocation densities  $\rho_f$, $\rho_{f}^*$ and $\tilde{\rho}_{m}$ were established in the following way.  The initial value of $\rho_f$ was computed by evaluating the above model for the as-annealed material to $\bar{\epsilon} = 0.14$ for the samples rolled at 0 and 90$^\circ$ to RD and  $\bar{\epsilon} = 0.16$ for the sample rolled at 45$^\circ$ to RD.  The value of $\rho_f$ at the end of these simulated tensile tests was taken as the starting value for the subsequent predictions of the rolled tensile samples (see table \ref{tab:initialrhos}).  As no aging was conducted (except for the short time spent at room temperature between rolling and tensile testing) the initial value of $\rho_f^*$ was taken to be zero.  This removes the impact of $\rho_f^*$ from these predictions.  In all cases at the end of the pre-deformation the model predicts that the material will have $\rho_m = \rho_m^{sat}$.  We adjust this, however, so that $\rho_m/\rho_m^{sat} < 1$ prior to re-loading to account for the fact that on change of strain path many of the dislocations that were active, will no longer be active (see e.g. \cite{Wilson1994}).  Thus, in this case we have only the initial value of $\rho_m$ as an adjustable parameter.  The values of $k_3$ and $k_4$ do not impact on the predictions because $\rho_f^*$ is taken to be zero.

For the rolled and aged samples we follow a similar process for predicting the value of $\rho_f$ at the pre-strain levels introduced by rolling. The best-fit values obtained here are reported in table \ref{tab:initialrhos}.  The initial values of $\rho_m$ established for the as rolled samples are retained assuming that the change in strain path has a larger impact than subsequent aging on the loss of mobiles dislocations.  The initial value of $\rho_f$ has been split into two components, an `unloaded' dislocation density ($\rho_f$) and a `loaded' dislocation density ($\rho_f^*$), where $\rho_f^*$ is taken to be much larger than $\rho_f$. A good fit could be found by assuming, for all three test directions, that the initial forest dislocation density in the as rolled material was partitioned 90\% to $\rho_f^*$ and 10\% remaining as $\rho_f$.  In the case of Al-Mg samples, it was necessary to account for the loss of dislocations due to recovery at the aging temperature.  Given the relatively low homologous temperature used for aging here, we envision (as a starting point) that recovery is small such that loss of dislocation density can be ignored. For aging at higher temperatures or longer times, this assumption would need to be revisited.   

\begin{table}[htbp]
 \centering
 \begin{tabular}{ccc}
  \toprule
  \textbf{Parameter} & \textbf{Value} & \textbf{Comment}  \\
  \midrule
  \multicolumn{3}{c}{Material Properties} \\
   \midrule
  $b$ & 0.205 nm & Burgers vector  \\
  $\mu$ & 80 GPa & Shear modulus  \\
  $M$ & 3 &  Taylor factor \\
  $\alpha$ & 0.35 & Saada constant  \\
  $m$ & 0.02 & Rate sensitivity  \\
  \midrule
  \multicolumn{3}{c}{Hardening Law Parameters (As-annealed/no strain aging)}   \\
  \midrule
  $\sigma_0$ & 175 MPa & \small{Yield stress (no strain aging)} \\
  $k_1$ & $2.0\times10^{8}$ m$^{-1}$ & \small{storage term} \\
  $k_2$ & 11.2 m$^{-1}$ & \small{annihilation rate} \\
  \midrule
  \multicolumn{3}{c}{Hardening Law Parameters (Pre-deformed $+$ Aged)}\\
  \midrule
  $\eta$ & 5 & generation rate of $\rho_m$ \\
  $k_{3}$ & $2.4\times10^{8} m^{-1}$ &  \\
  $k_{4}$ & 24 & \\
  $\rho_m^{sat}$ & $1\times10^{13}$ m$^{-2}$ & \\
  \bottomrule  
 \end{tabular}
 \caption{Parameters used in modelling the work hardening response using equations \ref{eqn:taylor}, \ref{eqn:rhom} and \ref{eqn:rhof2}. }
 \label{tab:params}
\end{table}

\begin{table}[htbp]
 \centering
 \begin{tabular}{cccc}
  \toprule
   Initial $\rho$ & \textbf{$0^\circ$} & \textbf{$45^\circ$} & $90^\circ$  \\
  \midrule
  \multicolumn{4}{c}{As rolled} \\
   \midrule
   $\rho_m$ & $\rho_m^{sat}$ &  $0.01\rho_m^{sat}$ & $0.01\rho_m^{sat}$ \\
   $\rho_f$ & $1.30 \times 10^{14}$, $m^{-2}$ & $1.35 \times 10^{14}$, $m^{-2}$ & $1.30 \times 10^{14}$, $m^{-2}$ \\
   $\rho_f^*$ & 0 & 0 & 0 \\
     \midrule
  \multicolumn{4}{c}{Rolled and Aged} \\
   \midrule
   $\rho_m$ & $0.01\rho_m^{sat}$ &  $0.01\rho_m^{sat}$ & $0.015\rho_m^{sat}$ \\
   Fraction $\rho_f$ & 0.1 & 0.1 & 0.1 \\
   Fraction $\rho_f^*$ & 0.9 & 0.9 & 0.9 \\
  \bottomrule  
 \end{tabular}
 \caption{Dislocation densities used in the prediction of the work hardening response after rolling and after rolling and aging. }
 \label{tab:initialrhos}
\end{table}

\begin{figure}[htbp]
\centering
\includegraphics[width=\figwidth\textwidth]{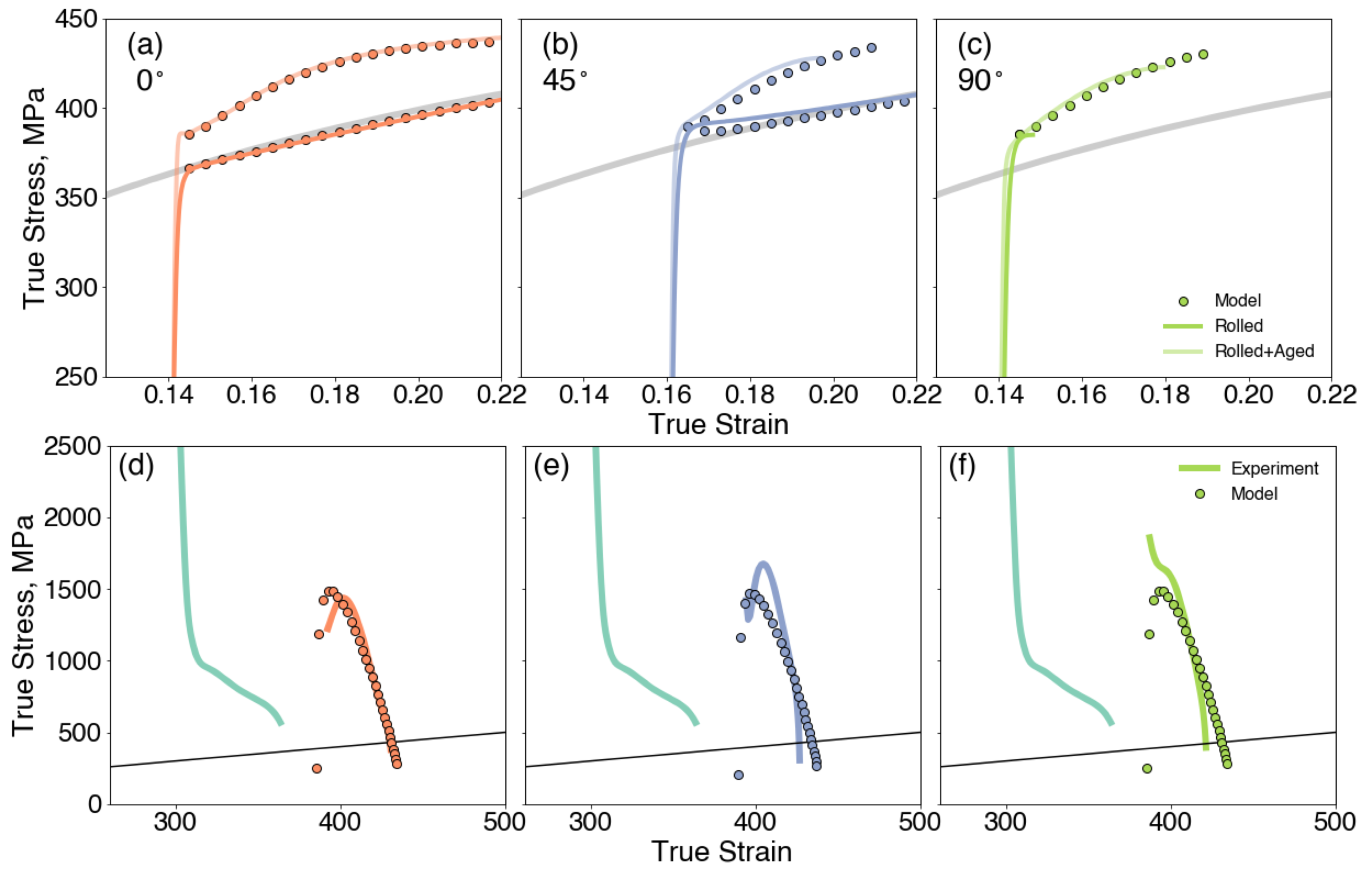}
\caption{a)-c) Stress-strain curves for samples after pre-deformation and pre-deformation and aging compared to model predictions ($o$). d)-f) shows the corresponding work hardening rates compared to model predictions.  The thick grey lines in each plot show the behaviour of the as-annealed material.}
\label{fig:modelresults}
\end{figure}

The resulting fits to the stress-strain behaviour of the rolled and rolled and aged samples using the parameters outlined in Tables \ref{tab:params} and \ref{tab:initialrhos} is shown in figure \ref{fig:modelresults}.  One can see that the model predicts the yield strength and work hardening behaviour of the samples tested at 0 and 45$^\circ$ to RD while predicting (using the same parameters) the initial yield strength of the sample tested at 90$^\circ$ to RD.  It can be seen that the initial transient in the work hardening behaviour following yielding for the sample tested at 45$^\circ$ to RD is captured by the reduced mobile dislocation density.

The model also does a very good job of predicting the yield strength and work hardening behaviour of the rolled and aged samples, including the initial transients in the near-yield portion of the stress strain curves from tests at 0 and 45$^\circ$ to RD.  The work hardening rate at high stress is somewhat over-predicted for the samples tested at 45 and 90$^\circ$ to RD.  This likely could be improved by refining the underlying model but only at the cost of increased complexity. 

\begin{figure}[htbp]
\centering
\includegraphics[width=\figwidth\textwidth]{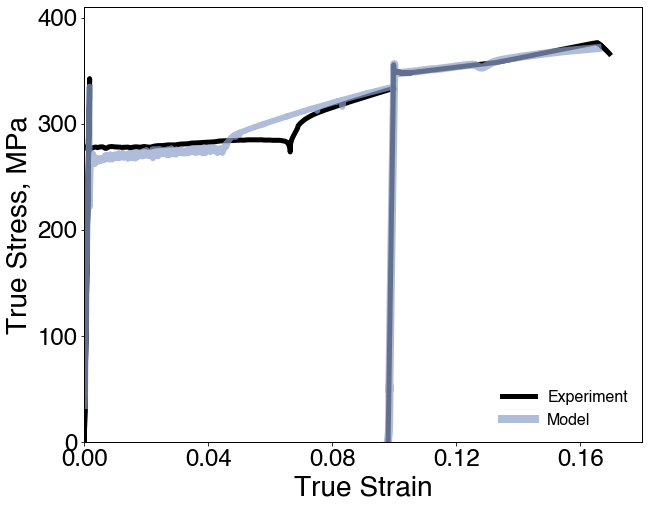}
\caption{Illustration of the use of the constitutive model developed here as a VUHARD routine within ABAQUS. The model has been applied to the data shown in figure \ref{fig:strainagekinetics}a, i.e. the as-annealed sample and a sample pre-strain in tension, aged at 150$^\circ$C for 10 min.}
\label{fig:modelFE}
\end{figure}

As a final illustration of the utility of this model, we return to the `classic' case of strain aging under monotonic loading (figure \ref{fig:strainagekinetics}), i.e. pre-deformation in tension followed by aging and re-loading in tension parallel to the prior loading direction. Figure \ref{fig:modelFE} shows the predictions made using the above model as a user defined hardening subroutine (VUHARD) within an ABAQUS finite element simulation.  For the as-annealed material, in order to predict the large upper yield point it was necessary to set the initial mobile dislocation density to a very small value ($\sim$ 1 m$^{-2}$).  This is in line with the interpretation \cite{Hahn1962, Gilman1959} that annealing and room temperature aging is sufficient to lock nearly all available mobile dislocations and sources.  The very high upper yield stress reflects the lack of mobile dislocations in equation \ref{eqn:taylor}.  The finite element predicted stress-strain curve under-predicts the duration of the L\"uder's plateau but this may, in part, be due to the fact that the strain rate dependency at the L\"uder's front was not accounted for, instead the macroscopic strain rate was used in equation \ref{eqn:taylor}.

In the case of prediction for aging a sample at 150$^\circ$C for 10 minutes, the model also does an excellent job of predicting the stress-strain response.  In this case, to obtain the observed results the fraction of forest dislocations converted to loaded forest dislocations ($\rho_f^*$) had to be taken to be lower than that for the rolled and aged samples (70\% in this case, compared to 90\% in the case of the rolled and aged samples, cf. table \ref{tab:initialrhos}).  In addition, the mobile dislocation density following aging had to be set to a much lower value ($1\times10^{-4}\rho_m^{sat}$) to give the small upper yield point and yield point elongation.  Again, the impact of the high strain rate at the L\"uder's front was not considered.   

It is interesting to reflect upon the difference in behaviour observed in the case of the tension-aged-tension and rolled-aged-tension (particularly 0$^\circ$ to RD) tested samples.  Clearly the minimal model applied here is unable to capture all of the complex physics involved in strain path changes and the results on the stress-strain behaviour, yet it is able to reproduce key aspects of the stress-stain response.  As the level of pre-strain (in rolling) is increased it may be expected that the model will have difficulty capturing the increasing load-path-change induced transients on loading \cite{Wilson1994}.  It is also not obvious why, in tension-aging-tension conditions, the dominant effect shifts from loaded forest dislocations ($\rho_f^*$) towards a reduced mobile dislocation density.  Finally, we anticipate that the model presented here will not account for effects that may arise from backstresses and kinematic hardening.  As mentioned in the introduction, measurements of the effects of aging when the loading direction is fully reversed (tension-aging-compression) show a lowering of the yield stress and an extended elasto-plastic transition \cite{Saada2013,Lai2018}. From a macroscopic perspective, a number of models have been developed which explicitly account for kinematic hardening via changes to the yield surface (see e.g. \cite{Manik2015,Qin2017} for examples of recent work).  Alternatively, microscopic models have been developed (e.g. \cite{Kitayama2013,Wen2016}) which explicitly build in an evolving back stress at the slip system level, based on classic concepts of back stress development due to inhomogeneous dislocation distributions \cite{Mughrabi1983}. Incorporating the additional effects induced by strain aging on such backstress evolution laws would require consideration of the relaxation of back stresses during strain aging as well as the evolution of a new set of back stresses on re-loading in the presence of a inhomogeneous distribution of solute atoms.  This is an interesting direction for future consideration.

\section{Conclusion}

A recently developed model for strain aging \cite{Medrano2020} has been adapted to predict the measured stress-strain response of a strain aged ULC steel subjected to a strain path change, this being more representative of typical application than the common monotonic pre-deformation in tension, aging, testing in tension procedure typically used to character strain aging.  It is found here that beyond changes to the yield stress, the work hardening response was changed significantly upon strain aging followed by a change in strain path. In particular, it was found that the initial work hardening rate at yield was significantly higher than that without strain aging. This was attributed, primarily, to forest dislocations `loaded' by carbon segregation.  These `loaded' dislocations are assumed to dynamically recover at a slower rate than non-loaded dislocations and therefore induce a higher rate of forest dislocation storage over the initial portion of the stress-strain curve.  As these `loaded' dislocations recover, the work hardening response evolves back towards that expected in the absence of strain hardening.  Testing this isotropic hardening model under a variety of other strain paths/loading conditions would be an ideal way to test the limits of the model and to identify conditions (like fully reversed strain path changes) where it is expected that the model will need to be improved.

\section{Acknowledgements}
The authors would like to acknowledge funding from the Canadian Natural Science and Engineering Research Council of Canada for funding and the provision of materials by ArcelorMittal Dofasco.

\bibliography{Thesis_ref.bib}

\end{document}